\documentclass[]{tMOP2e}

\usepackage{graphicx}
\usepackage{color}
\usepackage{subfigure}
\usepackage{mathrsfs}
\usepackage{amsmath}

\theoremstyle{plain}

\theoremstyle{remark}

\renewcommand{\epsilon}{\varepsilon}

\begin{document}
\title{Normalization of the wavefunction obtained from perturbation theory based on a matrix method}

\author{
\name{B.M. Villegas-Mart\'inez$^1$\thanks{$^\ast$Corresponding author. Email: bvillegas@inaoep.mx \vspace{6pt}}, H. M. Moya-Cessa$^1$ and F. Soto-Eguibar$^1$
\\\vspace{6pt} 
$^1$ Instituto Nacional de Astrof\'{\i}sica, \'Optica y Electr\'onica, Calle Luis Enrique Erro No. 1, Santa Mar\'{\i}a Tonantzintla,  Puebla, M\'exico 72840.}
}

\maketitle

\begin{abstract}
We present the derivation of the normalization constant for the perturbation matrix method recently proposed. The method is tested on the problem of a binary waveguide array for which an exact and an approximate solution are known. In our analysis, we show that to third order the normalized matrix method approximate solution gives results coinciding  with the exact known solution.
\end{abstract}

\section{Introduction}
In quantum mechanics the wavefunction plays an important role since it contains all the relevant information about the dynamical behavior of a physical quantum system \cite{1,2}. To determine the evolution of the wavefunction, one must solve the time-dependent Schr\"{o}dinger equation \cite{3}; in consequence, an analytic exact solution of this equation is of utmost importance to extract the whole appropriate information about a system \cite{4,5}. Nevertheless, the number of problems that can be solved analytically is limited \cite{6,7,8}; for example, the harmonic oscillator \cite{9}, the finite depth potential well \cite{10} and the hydrogen atom \cite{11}, among others. Under these circumstances, it is necessary to resort to methods, like perturbation theory, that allows to obtain an approximate solution \cite{12}; one of these methods is the Rayleigh-Schr\"{o}dinger perturbation theory, which has been widely used in many branches of physics with the main objective of finding solutions very close to the exact one \cite{13}. In spite of this, there are particular cases where the obtained results using the Rayleigh-Schr\"{o}dinger perturbation theory show serious problems of convergence \cite{14}; for instance, the quantum system of a charged particle hopping on an infinite linear chain driven by an electric field \cite{15}. This has prompted researchers to develop and implement new techniques to get better approximate solutions for the time-dependent Schr\"{o}dinger equation.
\\
An alternative perturbative approach, that we will focus along this work, is the Matrix Method \cite{16,17}.  {This new scheme, based on the implementation of triangular matrices, allows to solve approximately the time-dependent Schr\"odinger equation in an elegant and simple manner. The method has demonstrated that the corrections to the wavefunction and the energy can be contained in only one expression, unlike the standard perturbation theory where it is needed to calculate them in separated ways \cite{16,17}. Moreover, the Matrix Method may also be used when one may find a unitary evolution operator for the unperturbed Hamiltonian but it is not possible to find its eigenstates. }

 {On the other hand, its approximated solutions not only present conventional stationary terms, but also time dependent factors which allows us to know the temporal evolution of the corrections. A remarkable feature is that the general expression to compute them does not distinguish if the Hamiltonian is degenerate or not \cite{16,17}. Besides, the formalism offers an alternative to express the Dyson series in a matrix form. Even, an extension of this mathematical analysis has been developed to obtain approximative solutions to the Lindblad-type master equation \cite{18}. Therefore, the Matrix Method possess many attractive features that cannot be found in the conventional treatments of the perturbation theory. However, it is worth noting that the perturbed solutions of the Schr\"{o}dinger equation using the Matrix Method are not normalized, therefore, the main goal of this work is to find the normalization factor which gives us a complete perturbative description of the solutions.}
\\
The remainder of this work is organized as follows: In Section 2, we briefly review the Matrix Method described above. In Section 3, making use of the formalism of the Matrix Method, we show the development to obtain the general expression to compute the normalization constant to any order. In Section 4, we apply this approach to the particular problem of a binary waveguide array, which not only has a known exact analytic solution, but also has an approximate solution given by the small rotation method. Hence, in this particular problem, we shall show that the behavior of the field intensity distribution, given by the perturbed normalized result for the first three waveguides, has a very similar behavior with the known exact solution and has a much better accuracy than the solution obtained by the small rotation method. Finally, conclusions are given in Section 5.

\section{The Matrix Method}
The Matrix Method \cite{16,17,18} arises from the formal solution of the time-dependent Schr\"{o}dinger equation $\left|\psi(t)\right \rangle=e^{-i t \hat{H}} \left|\psi(0) \right \rangle$, with the complete Hamiltonian $  \hat{H}= \hat H_0 +\lambda \hat{H_p}$ divided into an unperturbed part $\hat H_0$ and a perturbed part  $\hat{H_p}$, being $\lambda $ a perturbation parameter. The formal solution of the Schr\"{o}dinger equation can be expanded in a Taylor series and sorted in powers of $\lambda$; for example, the expansion up to first-order is
\begin{equation} \label{1}
\left|\psi(t) \right \rangle=  \left[ e^{-i \hat H_0 t} + \lambda \sum\limits_{n=1}^{\infty} \frac{(-it)^n}{n!}  \sum\limits_{k=0}^{n-1} \hat H_0^{n-1-k} \hat{H_p} \hat{H_0}^{k}\right]     \left| \psi(0)\right \rangle.
\end{equation}
The key to simplify and get a solution of the above is to consider the triangular matrix
\begin{equation} \label{2}
M=
 \begin{pmatrix}
 \hat {H_0} & \hat{H_p}  \\
  0 & \hat H_0  \\
  \end{pmatrix},
\end{equation}
whose diagonal elements are conformed by the unperturbed part of the Hamiltonian and the superior triangle by the perturbation. One can find that if we multiply the matrix $M$ by itself $n$-times, its upper element will contain exactly the same products of $\hat H_0$ and $\hat{H_p}$ defined within summation in Eq.\eqref{1}. In other words, the matrix element $M_{1,2}$ will give us the first order correction; based on this consideration, we arrive to
\begin{equation} \label{3}
\left|\psi(t) \right \rangle= \left[ e^{-i \hat H_0 t} + \lambda (e^{-i M t})_{1,2}\right]  \left| \psi(0)\right \rangle,
\end{equation}
where the first split part corresponds to the zero-order term and the other part refers to the first order correction. The above relation can be rewritten as
\begin{equation} \label{4}
\left|\psi(t) \right \rangle=  \left| \psi^{(0)} \right \rangle + \lambda \left( \left| \psi^{P} \right \rangle\right)_{1,2},
\end{equation}
where $\left| \psi^{P} \right \rangle$ is a matrix defined as
\begin{equation} \label{5}
\left| \psi^{P} \right \rangle=
 \begin{pmatrix}
 \left| \psi_{1,1} \right \rangle & \left| \psi_{1,2} \right \rangle  \\
  \left| \psi_{2,1} \right \rangle & \left| \psi_{2,2} \right \rangle \\
  \end{pmatrix}.
\end{equation}
The solution to first order can be determined if we derived the equations \eqref{3} and \eqref{4} with respect to time, equate the corresponding coefficients of $\lambda$ and perform the algebraic steps outlined in \cite{16,17}; we obtain
\begin{equation} \label{6}
\left|\psi_{1,2} \right \rangle = -i e^{-i \hat H_0 t}\left[\int\limits_0^t  e^{i \hat H_0 t_1} \hat{H_{p}}
  e^{-i \hat H_0 t_1}  dt_1 \right]\left|\psi(0) \right \rangle.
 \end{equation}
All the information for the second order correction will be enclosed in the element $M_{1,3}$ of a newly defined $3 \times 3$ triangular matrix $M$, completely similar to \eqref{2}. Thus, the Matrix Method allows to transform the Taylor series of the formal solution of the time-dependent Schr\"{o}dinger equation in a power series of the matrix $M$, which can be handled easily. Likewise, this process allows us to find any $k$th-order correction in a simple and straightforward way through the following relation \cite{16,17}
\begin{equation}\label{7}
\left|\psi(t) \right \rangle = \left| \psi^{(0)} \right \rangle + \sum\limits_{j = 1}^{k} \lambda^j \left( \left| \psi^{P} \right \rangle\right)_{1,j+1},
\end{equation}
with
\begin{equation} \label{8}
\left| \psi^{P} \right \rangle=
\begin{pmatrix}
     \left|\psi_{1,1} \right \rangle & \hdots & \left|\psi_{1,j+1} \right \rangle \\
    \vdots & \ddots & \vdots \\
    \left|\psi_{j+1,1}\right \rangle & \hdots & \left|\psi_{j+1,j+1} \right \rangle \end{pmatrix},
\end{equation}
being the matrix element $\left|\psi_{1,j+1} \right \rangle$ the relevant solution we are looking for, which is expressed in the form
\begin{equation}\label{9}
	\left|\psi_{1,j+1} \right \rangle = -i e^{-i \hat H_0 t}\int\limits_0^t e^{i \hat H_0 t_1} \hat{H_p}
	\Biggl[ -i e^{-i \hat H_0 t_1} \int\limits_0^{t_1 } 
	e^{i \hat H_0 t_2}\hat{H_p} \Biggl[-i e^{-i \hat H_0 t_2} \int\limits_0^{t_2 }
	.....dt_3  \Biggr] dt_2\left|\psi(0) \right \rangle \Biggr]  dt_1  .
\end{equation}
This time-ordered series, restricted to the interval $ [0, t] $, is the fundamental piece to calculate the different correction terms; furthermore, we should point out that the relationship \eqref{9} is the mathematical representation of the Dyson series \cite{19,20}. { Although this latter expression only applies for weak perturbations, its strong perturbation counterpart ($\lambda \to \infty $) can be derived in a straightforward way interchanging the unperturbed Hamiltonian $\hat{H_0}$ with the perturbation $\hat{H_p}$ and rescaling the time as $\tau=\lambda t$. This duality on the Matrix Method gives us the possibility to analyze the solution of a quantum system in both regimes of the perturbative parameter. However, in this work, we are limited to weak perturbations, as the example in Section 4.}

\section{Normalization constant}
In the previous section, we appreciated that the approximated solution of the Schr\"{o}dinger equation can be written as a power series of the perturbative parameter $\lambda$, along with the element $\left|\psi_{1,j+1} \right \rangle$ of the perturbed matrix. It is appropriate to mention that the expression \eqref{7} is not normalized and it is convenient to get a normalization factor $N_k$ that preserves its norm at any order. Then, let us define the next normalized solution
\begin{equation} \label{10}
\left|\Psi(t) \right \rangle = N_k \left(   \left| \psi^{(0)} \right \rangle + \sum\limits_{j = 1}^{k} \lambda^j \left| \psi_{1,j+1} \right \rangle  \right),
\end{equation}
where the corresponding value of $N_k$ may be easily determined by the normalization condition $\left\langle \Psi(t) | \Psi(t) \right\rangle= 1$ and can be expressed as follows
\begin{equation} \label{11}
N_k=\left[ 1+ 2 \sum\limits_{j = 1}^{k} \lambda^j \Re \left( \left\langle \psi^{(0)} |  \psi_{1,j+1} \right \rangle \right)+  \sum\limits_{m,j = 1}^{k}\lambda^{m+j} \left\langle \psi_{1,m+1} | \psi_{1,j+1} \right\rangle \right]^{-\frac{1}{2}},
\end{equation}
being the first contribution due to $ \left\langle \psi^{(0)} | \psi^{(0)} \right \rangle $,  whereas the second one arises from two single finite sums, one referent to the inner product of the zero-order term with the $j$th-order correction $\left\langle \psi^{(0)} |  \psi_{1,j+1} \right \rangle$, and the other respect to its complex conjugate $\left\langle \psi_{1,j+1} | \psi^{(0)} \right \rangle$; as consequence, a purely real contribution is obtained of the sum of both over all $k$. The last part of the above equation is merely handled if we run $m$ and $j$ from $1$ to $k$, such as presented in Table \ref{tab:1}.
\begin{table}
\centering{
\resizebox{12cm}{!} {
\begin{tabular*}{14.5 cm}{ l| c c c c c @{\extracolsep{\fill}} r| }
 $m$/$j$&1&2&3&$\ldots$&$k$ \\
\hline\\
1 & $\lambda^2 \left\langle \psi_{1,2} | \psi_{1,2} \right\rangle$  &  $\lambda^3 \left\langle \psi_{1,2} | \psi_{1,3} \right\rangle$ &  $\lambda^4 \left\langle \psi_{1,2} | \psi_{1,4} \right\rangle$ &$\ldots$&  $\lambda^{1+k} \left\langle \psi_{1,2} | \psi_{1,k+1} \right\rangle$ \\
2 & $\lambda^3 \left\langle \psi_{1,3} | \psi_{1,2} \right\rangle$  &  $\lambda^4 \left\langle \psi_{1,3} | \psi_{1,3} \right\rangle$ &  $\lambda^5 \left\langle \psi_{1,3} | \psi_{1,4} \right\rangle$  &$\ldots$&  $\lambda^{2+k} \left\langle \psi_{1,3} | \psi_{1,k+1} \right\rangle$\\
3 & $\lambda^4 \left\langle \psi_{1,4} | \psi_{1,2} \right\rangle$  &  $\lambda^5 \left\langle \psi_{1,4} | \psi_{1,3} \right\rangle$ &  $\lambda^6 \left\langle \psi_{1,4} | \psi_{1,4} \right\rangle$  &$\ldots$&  $\lambda^{3+k} \left\langle \psi_{1,4} | \psi_{1,k+1} \right\rangle$\\
$\vdots$ & $\vdots$ & $\vdots$ & $\vdots$ &   & $\vdots$ \\
$\vdots$ & $\vdots$ & $\vdots$ & $\vdots$ & $\ddots$ & $\vdots$ \\
$ k $ & $\lambda^{1+k} \left\langle \psi_{1,k+1} | \psi_{1,2} \right\rangle$ & $\lambda^{2+k} \left\langle \psi_{1,k+1} | \psi_{1,3} \right\rangle$ & $\lambda^{3+k} \left\langle \psi_{1,k+1} | \psi_{1,4} \right\rangle$ &  $\ldots$ & $\lambda^{2k} \left\langle \psi_{1,k+1} | \psi_{1,k+1} \right\rangle$
\end{tabular*}}}
\caption{This table displays the different terms of the double summation contained in the second term of Eq. \eqref{11}, when $m$ and $j$ run from 1 to $k$}
\label{tab:1}
\end{table}
It is easy to see that the double summation in  \eqref{11} can be split in two parts, one where $m=j$ and which contains all diagonal terms, and the other for the off-diagonal terms represented in a double sum of the real part of $ \left\langle \psi_{1,m+1} | \psi_{1,j+1} \right\rangle$, i.e.
\begin{equation}\label{12} 
 \sum\limits_{m,j = 1}^{k}\lambda^{m+j} \left\langle \psi_{1,m+1} | \psi_{1,j+1} \right\rangle = \sum\limits_{n = 1}^{k}\lambda^{2n} \left\langle \psi_{1,n+1} | \psi_{1,n+1} \right\rangle 
 + 2 \sum_{\substack{n=1\\ k>1}}^{k-1} \sum\limits_{m = n+1}^{k} \lambda^{n+m}\Re \left( \left\langle \psi_{1,n+1} | \psi_{1,m+1} \right\rangle \right);
\end{equation}
applying the change of variable $m=p-n$ and replacing in the Eq.\eqref{11}, we arrive to
\begin{equation}\label{13}
\begin{split}
N_k = \left[ 1 + 2 \sum\limits_{j = 1}^{k} \lambda^j \Re \left( \left\langle \psi^{(0)} |  \psi_{1,j+1} \right \rangle \right)   + \sum\limits_{n = 1}^{k}\lambda^{2n} \left\langle \psi_{1,n+1} | \psi_{1,n+1} \right\rangle
 \right. \\  \left.
+2 \sum_{\substack{n=1\\ k>1}}^{k-1} \sum\limits_{p = 2n+1}^{n+k} \lambda^{p}\Re \left( \left\langle \psi_{1,n+1} | \psi_{1,p-n+1} \right\rangle \right) \right]^{-\frac{1}{2}},
\end{split}
\end{equation}
which is the normalization constant for the approximate analytical solution of the Schr\"{o}dinger equation defined in Eq.\eqref{10}. In principle, the inclusion of the factor $N_k$ in our calculations can give a fairly good approximation to the solution without convergence difficulties.  {An important remark on the proposed normalization procedure is that we have not invoked the usual intermediate normalization used in the standard perturbation theory, i.e. the imposition $\left\langle \psi^{(0)} | \psi_{1,n+1} \right\rangle=0$ for all $\lambda$. Such condition does not apply in our case as can be seen clearly in Appendix B, where it is shown that the inner product of the zero-order correction with the first two correction terms are different from zero. In particular, these complex inner products have non-zero imaginary parts which should not be neglected if the called intermediate normalization is applied; for this reason, we have adopted other procedure to obtain the factor $ N_{k}$, which ensures real values at any power of $\lambda$.}

\section{The Binary waveguide array as an example}
\begin{figure} [h]
  \centering
  \begin{tabular}{@{}c@{}} 
    \includegraphics[width=0.327 \linewidth]{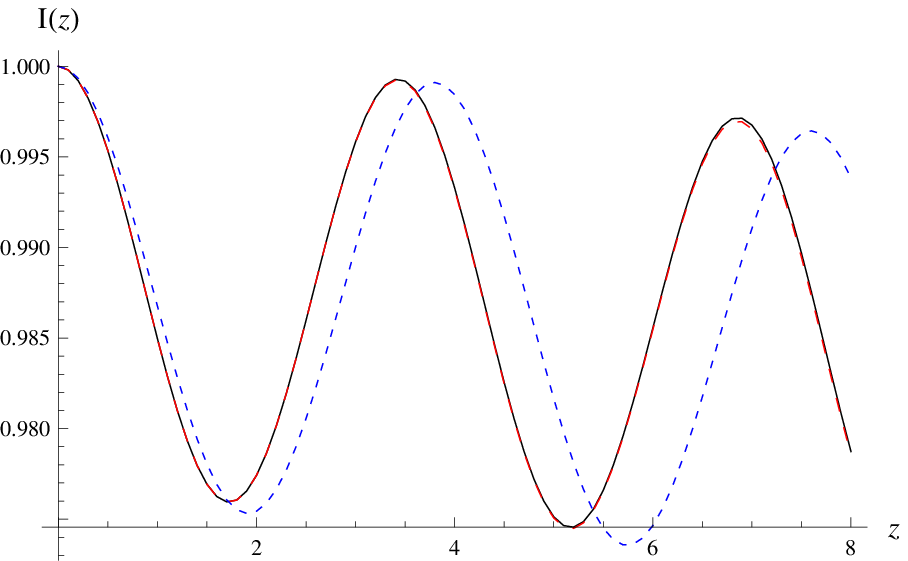} \\ 
     \small (a) Guide 1
  \end{tabular}
   \begin{tabular}{@{}c@{}} 
    \includegraphics[width=0.327 \linewidth]{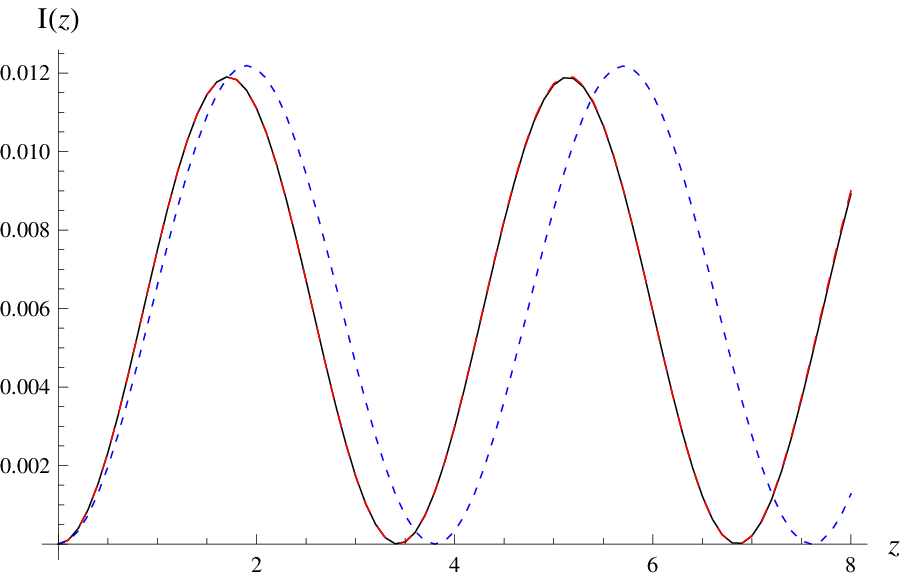} \\
    \small (b) Guide 2
  \end{tabular}
 \begin{tabular}{@{}c@{}}
    \includegraphics[width=0.327 \linewidth]{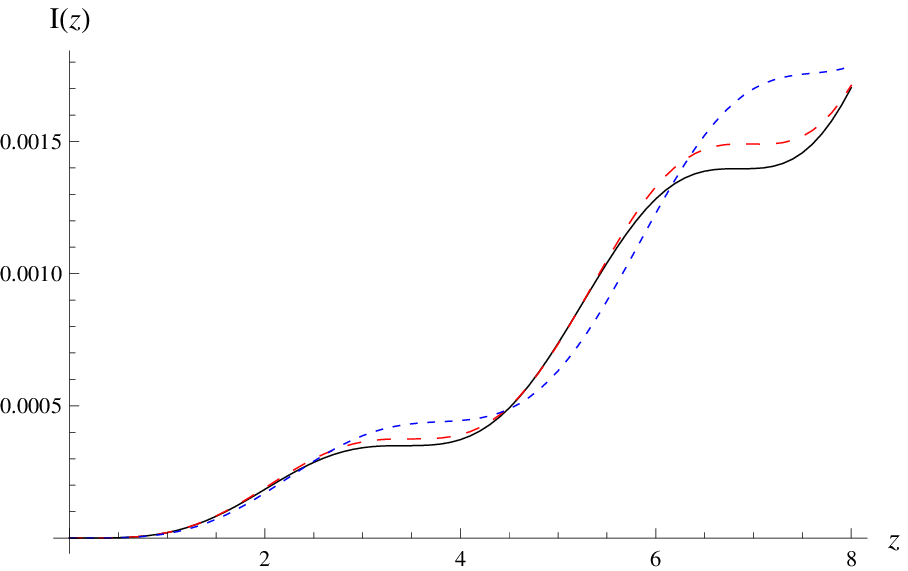} \\
    \small (c) Guide 3
  \end{tabular}
\caption{Field intensity versus propagation distance $z$ using the exact solution (solid line), the third-order solution (red dashed line) and the small rotation method solution (blue dashed line), with $\alpha=0.1$  and $\omega=0.9$, for the first three guides.} \label{fig:1}
\end{figure}

\begin{figure} 
  \centering
  \begin{tabular}{@{}c@{}} 
    \includegraphics[width=0.327 \linewidth]{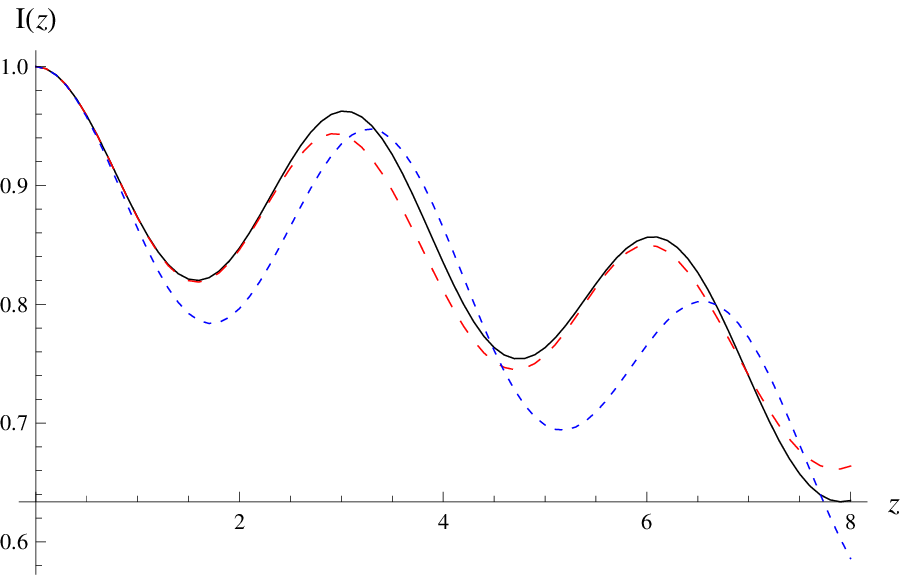} \\ 
     \small (a) Guide 1
  \end{tabular}
   \begin{tabular}{@{}c@{}} 
    \includegraphics[width=0.327 \linewidth]{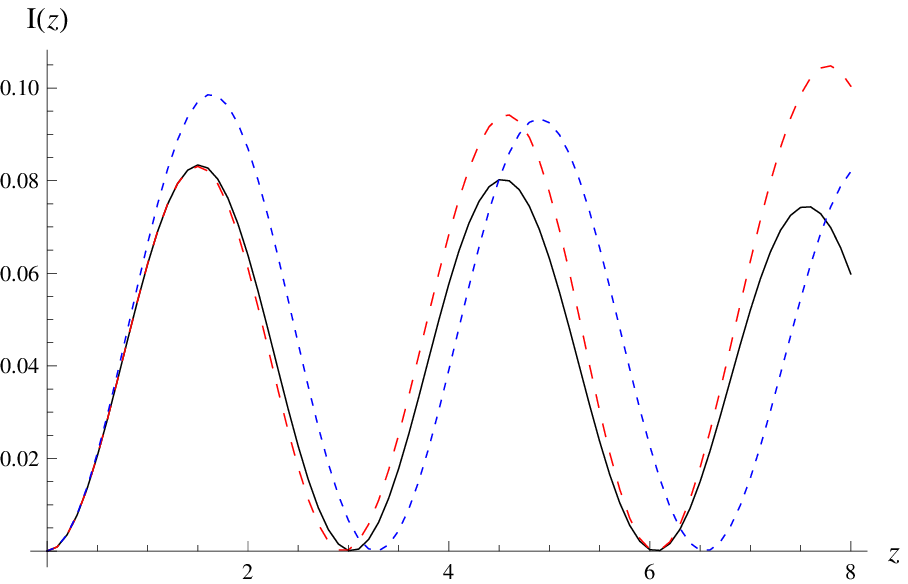} \\
    \small (b) Guide 2
  \end{tabular}
 \begin{tabular}{@{}c@{}}
    \includegraphics[width=0.327 \linewidth]{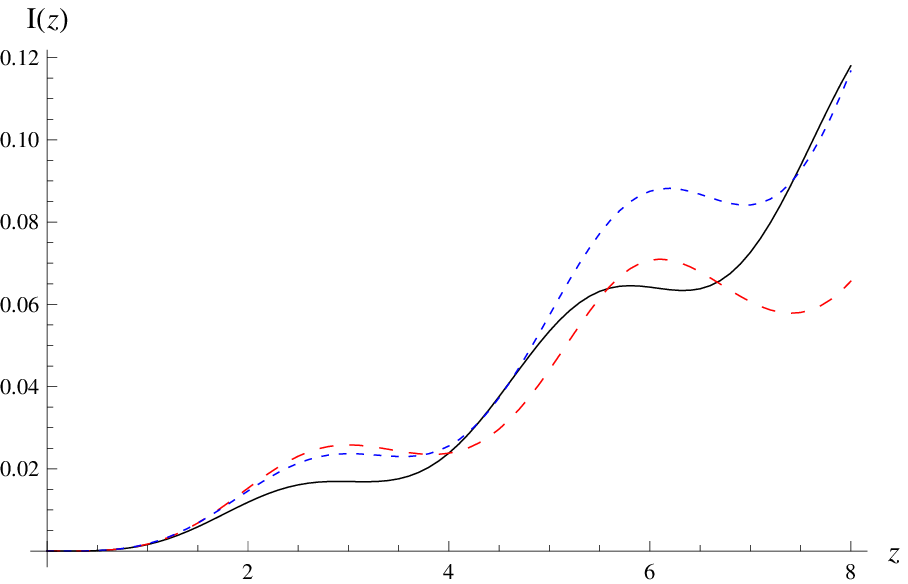} \\ 
    \small (c) Guide 3
  \end{tabular}
\caption{Field intensity versus propagation distance $z$ using the exact solution (solid line), third-order solution (red dashed line) and the small rotation method solution (blue dashed line), with $\alpha=0.3$  and $\omega=0.9 $, for the first three guides.} \label{fig:2}
\end{figure}

In order to test the accuracy of our perturbative method, we apply it to a problem whose exact solution is known; in this case, a  waveguide array. These optical structures have demonstrated its potential to emulate some particular problems related with quantum mechanics, such as optical Bloch oscillations \cite{21}, discrete spatial solitons \cite{22}, quantum walks \cite{23}, discrete Fourier transforms \cite{Leija} and parity time-symmetry \cite{24}, to name a few. In particular, the linear behavior of light propagation over this kind of waveguide arrangement is usually governed by the infinite system of differential equations
\begin{equation} \label{14}
i\frac{d \mathscr{E}_{n}}{d z}= \omega \left(-1 \right)^n \mathscr{E}_{n} + \alpha \left(\mathscr{E}_{n+1}+ \mathscr{E}_{n-1} \right),      \qquad n=-\infty,...,-2,-1,0,1,2,...,\infty,
\end{equation}
where $\mathscr{E}_{n}$ represents the amplitudes of the light field confined in the $n$th waveguide, $z$ the longitudinal propagation distance, $2 \omega$ the mismatch propagation constant and $\alpha$ the hopping rate between two adjacent waveguides. Physically, equation \eqref{14} describes the effective evanescent field coupling between the nearest-neighbor waveguide interactions. Moreover, it has been demonstrated \cite{25} that this system can be associated with a Schr\"{o}dinger-type equation
\begin{equation}\label{15}
i \frac{d \left|\psi(z)\right \rangle}{d z}= \hat{H}\left|\psi(z)\right \rangle,
\end{equation}
where $\hat{H}=\omega \left(-1\right)^{\hat{n}} + \alpha \left(\hat{V}+\hat{V}^\dagger  \right)$, being $\hat{n}$ the number operator, $\left(-1\right)^{\hat{n}}$ the parity operator and $\hat V$ and $\hat{V}^\dagger$ the ladder operators defined as
\begin{equation} \label{16}
\hat V= \sum_{n=- \infty}^\infty \left|n \right \rangle \left\langle n+1 \right| , \quad \hat{V}^\dagger= \sum_{n=-\infty}^\infty \left|n+1 \right \rangle \left\langle n \right|.
\end{equation}
Note that if the solution is written in terms of the orthonormal (Wannier) states \cite{Kenkre} as $\left|\psi(z)\right \rangle= \displaystyle\sum_{n=-\infty}^\infty \mathscr{E}_ {n} (z) \left|n \right \rangle$, and if this proposal is substituted into Eq.\eqref{15}, the infinite system given by Eq.\eqref{14} is recovered. In fact, it has been shown \cite{26}, that the exact solution of this system is
\begin{equation} \label{17}
\mathscr{E}_{n}(z)= \frac{1}{\pi} \int\limits_0^\pi {\cos \left( n\phi\right)  \left\lbrace \cos[\Omega(\phi)z] -i[2\alpha \cos\phi + (-1)^n \omega]\frac{\sin[\Omega(\phi) z]}{\Omega(\phi)} \right\rbrace  d\phi },
\end{equation}
with
\begin{equation} \label{18}
\Omega(\phi)=\sqrt{\omega^2 + 4 \alpha^2 \cos^2\phi }.
\end{equation}
Despite the above equation represents the exact solution for the amplitude of the light field, $\mathscr{E}_{n}= \left\langle n |\psi(z) \right \rangle $, a substantial alternative approximated solution is derived in \cite{26}; under the condition  $\alpha \ll \omega $ and performing the unitary transformation $ \tiny {\hat{R} =\exp{\left[ \frac{\alpha}{2\omega}\left( -1\right)^{\hat{n}} \left(\hat{V}+\hat{V}^\dagger  \right)\right] }}$, the following solution is found,
\begin{equation} \label{19}
\mathscr{E}_{n}=\left(-1 \right)^\frac{n(n-1)}{2} \sum\limits_{r =- \infty}^{\infty} \sum\limits_{s =- \infty}^{\infty} \left(-1 \right)^{s r}  e^{-i \left(-1 \right)^s \left(\frac{\omega^2+\alpha^2}{\omega} \right)z}i^{r}  
 \times J_r\left(\frac{\alpha^2}{\omega} z \right) J_s\left(\frac{\alpha}{\omega} z \right) J_{n+2r+s}\left(\frac{\alpha}{\omega} z \right),
\end{equation}
where $J_n(z)$ are the Bessel functions of the first kind \cite{25}; in fact, the unitary transformation $ \small{\hat{R}}$ constitutes the small rotation approximation \cite{27}.\\
Let us now solve the Schr\"{o}dinger-type equation with the formalism presented in Section 2. In the first place, we need to regard the variable $ z $ as the time; this is intuitively reasonable if we want to describe the optical field propagation on the waveguide array in evolutionary terms. Under this assumption, and considering that $\omega \left(-1\right)^{\hat{n}} $ is the unperturbed part, $\hat{V}+\hat{V}^\dagger $ is the perturbation and $\alpha $ as the perturbation parameter, one readily obtains from Eq. $\eqref{9}$,
\begin{equation}\label{20}
\left|\psi_{1,2} \right \rangle = -i \frac{\sin(\omega z) }{\omega} \left( \hat{V} + \hat{V}^\dagger  \right) \left|m \right \rangle.
\end{equation}
Since the problem is linear, we have considered the initial condition $\left|\psi(0)\right \rangle= \left|m\right \rangle $ which corresponds to a single excitation in the $m$th guide. The action of the ladder operators on this state leads to the first-order correction of the solution as follows
\begin{equation}\label{21}
\left|\Psi(z) \right \rangle  =  e^{-i \omega (-1)^{m} z} N_1 \left|m \right \rangle -i \frac{\alpha\sin(\omega z)}{\omega}  N_1 \left( \left|m-1 \right \rangle + \left|m+1 \right \rangle  \right),
 \end{equation}
with
\begin{equation} \label{22}
N_1= \left\lbrace  1+ 2 \left[  \frac{\alpha \sin(\omega z)}{\omega}\right]^2 \right\rbrace ^{-1/2}.
\end{equation}
All these derivations are offered in detail in the Appendix A. Now, if we consider the case in which the light field is launched into the first site of the waveguide array, i.e. $\left|m \right \rangle= \left|0 \right \rangle $ and $\mathscr{E}_{n}\left( z\right) = \left\langle n |\Psi(z) \right \rangle$, we get
\begin{equation} \label{23}
\mathscr{E}_n(z) =  e^{-i \omega z} N_1 \delta_{n,0} -i \frac{\alpha \sin(\omega z)}{\omega} N_1 \left( \delta_{n,-1} + \delta_{n,1} \right).
\end{equation}
Note that this equation describes the propagation of the electromagnetic field either towards the left side or towards the right side of the photonic waveguide array, but since this array is symmetric and infinite, we can simplify the analysis considering only the positive values of $n$. Therefore, the first-order solution is reduced to
\begin{equation}\label{24}
\mathscr{E}_n(z) =  e^{-i \omega z} N_1 \delta_{n,0} -i \frac{\alpha\sin(\omega z)}{\omega} N_1 \delta_{n,1},
\end{equation}
and satisfies the same initial conditions as those reported in the literature \cite{26}.\\
The second order term can be calculated using again the Eq.$\eqref{9}$,
\begin{equation} \label{25}
\left|\psi_{1,3} \right \rangle = i \frac{(-1)^m  \cos(\omega z)}{2 \omega^2} A(z)\left( \hat{V} + \hat{V}^\dagger \right)^2 \left|m \right \rangle,
\end{equation}
where the function $A(z)$ is defined by
\begin{equation}\label{26}
A(z)=\tan \left( \omega z \right)  \left[1+ iz\omega (-1)^m  \right]-z\omega.
\end{equation}
Thus, we can write the second-order correction as
\begin{align} \label{27}
\left|\Psi(z) \right \rangle  =& \cos\left( {\omega z}\right)  \left\{1 + i (-1)^m \left[ \frac{\alpha^2}{\omega^2} A(z)-\tan\left( {\omega z}\right)  \right] \right\} N_2 \left|m \right \rangle  - i \frac{\alpha\sin\left( \omega z\right)  }{\omega} N_2 \left( \left|m-1 \right \rangle + \left|m+1 \right \rangle  \right)
\nonumber \\ 
& + i \frac{ \alpha^2 (-1)^m A(z)  \cos\left(\omega z\right)    }{2 \omega^2} N_2 \left( \left|m-2 \right \rangle + \left|m+2 \right \rangle \right),
\end{align}
with its corresponding normalization constant
\begin{equation}\label{28}
N_2=\left\{ 1+\frac{3}{2} \left[ \frac{\alpha^2 \cos\left( \omega z\right)  \, || A(z)|| }{\omega^2} \right]^2 \right\}^{-1/2}.
\end{equation}
Now, assuming that the light is injected into the first guide, $m=0$, and considering the symmetry of the photonic array, we arrive to the reduced form of the field
\begin{align}\label{29}
\mathscr{E}_n (z) =& \cos(\omega z) \left\lbrace1 + i  \left[ \frac{\alpha^2}{\omega^2} A(z)-\tan(\omega z) \right]  \right\rbrace N_2 \delta_{n,0} -i \frac{\alpha\sin(\omega z)}{\omega}N_2 \delta_{n,1} \nonumber \\
& + i \frac{ \alpha^2 A(z) \cos(\omega z)  }{2 \omega^2} N_2 \delta_{n,2}.
\end{align}
The third-order term can be easily derived repeating the same steps used above for the first two order corrections,
\begin{equation}\label{30}
\left|\psi_{1,4} \right \rangle =\frac{ i}{2 \omega^3} \cos(\omega z) \left[ \tan(\omega z) - z\omega \right] \left( \hat{V} + \hat{V}^\dagger \right)^3 \left|m \right \rangle,
\end{equation}
and the complete solution is given by
\begin{align} \label{31}
\left|\Psi(z) \right \rangle &= \left\lbrace e^{-i\omega (-1)^m z} - \frac{\alpha^2 }{\omega^2} (-1)^m \left[
 z\omega (-1)^m \sin(\omega z) -i B(z) \right] \right\rbrace  N_3 \left|m \right \rangle \nonumber \\
& + i  \left[  \frac{3 \alpha^3 B(z)}{2\omega^3 } -\frac{\alpha \sin(\omega z) }{\omega} \right]  N_3 \left( \left|m-1 \right \rangle + \left|m+1 \right \rangle \right) \nonumber \\ &
-  \frac{\alpha^2 }{2 \omega^2} (-1)^m \left[ z\omega (-1)^m \sin(\omega z)  - i B(z)  \right] N_3 \left( \left|m-2 \right \rangle + \left|m+2 \right \rangle \right)
 \nonumber  \\ &
+ \frac{ i\alpha^3}{2 \omega^3} B(z) N_3  \left( \left|m-3 \right \rangle + \left|m+3 \right \rangle \right),
\end{align}
with
\begin{equation}\label{32}
B(z)=\cos(\omega z) \left[ \tan(\omega z) -z\omega \right]
\end{equation}
and
\begin{equation}\label{33}
N_3=\left\lbrace 1+ \frac{3}{2} \left(\frac{\alpha}{\omega}   \right)^4 \left[  B(z)^2 -4 B(z) \sin(\omega z) + z^2 {\omega}^2 \sin^2(\omega z) \right]+ 5 \left( \frac{\alpha}{\omega} \right)^6 B(z)^2\right\rbrace^{-1/2}.
\end{equation}
In this case, the third-order solution for the amplitude of the electric field on the waveguide array is given by
\begin{align} \label{34}
\mathscr{E}_n(z) =& \left\lbrace e^{-i\omega z}  - \frac{\alpha^2 }{\omega^2} \left[
 z\omega \sin(\omega z) -i B(z) \right] \right\rbrace N_3 \delta_{n,0} 
 + i  \left[  \frac{3 \alpha^3 B(z)}{2\omega^3 } -\frac{\alpha \sin(\omega z) }{\omega} \right]  N_3 \delta_{n,1} \nonumber \\
& - \frac{\alpha^2 }{2 \omega^2} \left[ z\omega  \sin(\omega z)  - i B(z)  \right]N_3 \delta_{n,2} 
+ \frac{ i\alpha^3}{2 \omega^3} B(z)  N_3  \delta_{n,3}.
\end{align}
 {In order to illustrate the high degree of accuracy that can be obtained with the third-order correction for the amplitude of the electrical field, the numerical comparison of this perturbative solution} with the exact solution and with the small rotation solution is given in Figs. \ref{fig:1} and \ref{fig:2}, using the parameters $\omega=0.9$ and $\boldsymbol{n=0,1,2}$. In these figures, we present the intensity distribution $I(z)=|\mathscr{E}_n|^2 $ for the first three guides considering two values of the perturbation parameter, $\alpha=0.1 $ and $\alpha=0.3 $. It is noteworthy that for $\alpha=0.1 $ the approximated solution converges to the exact solution uniformly even with a large propagation distance; for $\alpha=0.3$ both solutions are very similar only for short distances, but we still obtain a good approximation. Moreover, for these two values of $\alpha$, the third-order correction shows to be better than the small rotation approximation.


\section{Conclusions}

In summary, we have successfully obtained the normalization constant of Eq.\eqref{10} which complements the theoretical analysis of the Matrix Method. The perturbative solutions of the equations that describe the binary waveguide array, obtained applying this method, are highly accurate. Furthermore, it is shown that the third-order approximated solution matches exactly the known exact solution, not only for small values of the perturbative parameter $\alpha$, but also for large values; in fact, the real measure of the perturbation is the product $\alpha z$. On the other hand, it becomes evident the improvement of this method with respect to the reported results using the small rotation method. Therefore, the assessment of higher-order terms can give us a reliable solution into the system described here.

\section*{Acknowledgments}
 B.M. Villegas-Mart\'inez thanks the support given by the National Council on Science and Technology (CONACYT).

\appendix
\section{\null}

From Eq. $\eqref{10}$, we can write the analytic solution to first-order as
\begin{equation}\label{35}
\left|\Psi(t) \right \rangle = N_1 \left| \psi^{(0)} \right \rangle + \alpha N_1  \left| \psi_{1,2} \right \rangle.
\end{equation}
The zero-order solution is trivial, $\left| \psi^{(0)} \right \rangle= e^{-i \omega (-1)^{\hat{n}} z} \left|m \right \rangle $. The first-order correction, $\left| \psi_{1,2} \right \rangle $, requires the use of Eq.$\eqref{9}$,  together with the conditions established in Sec.4, and gives
\begin{equation} \label{36}
\left|\psi_{1,2} \right \rangle =-i e^{-i \omega (-1)^{\hat{n}} z}\int\limits_0^z    e^{i \omega (-1)^{\hat{n}} z_1} \left( \hat{V} + \hat{V}^\dagger  \right) e^{-i \omega (-1)^{\hat{n}} z_1}  \left|m \right \rangle dz_1 .
\end{equation}
We expand in Taylor series the product of operators inside the integral as
\begin{equation*} \label{37}
 e^{i \omega (-1)^{\hat{n}} z_1} \left( \hat{V} + \hat{V}^\dagger \right) e^{-i \omega (-1)^{\hat{n}} z_1} = \sum\limits_{l,r = 1}^{\infty} \frac{\left(-1\right)^r \left(i \omega z_1 \right)^{l+r}}{l! r!} \left(-1 \right)^{l \hat{n}} \left( \hat{V} + \hat{V}^\dagger \right) \left(-1 \right)^{r \hat{n}} ,
\end{equation*}
but
\begin{align*}
\left(-1 \right)^{l \hat{n}} \left(\hat{V} + \hat{V}^\dagger \right) \left(-1 \right)^{r \hat{n}}\left|m \right \rangle &=  \left(-1 \right)^{r m} \left(-1 \right)^{l \hat{n}} \left( \left|m-1 \right \rangle + \left|m+1 \right \rangle  \right) \\
&=\left(-1 \right)^{l} \left(-1 \right)^{(r+l)m} \left( \hat{V} + \hat{V}^\dagger \right) \left|m \right \rangle,
\end{align*}
then
\begin{align*}
e^{i \omega (-1)^{\hat{n}} z_1} \left( \hat{V} + \hat{V}^\dagger \right) e^{-i \omega (-1)^{\hat{n}} z_1} & = \sum\limits_{l,r = 1}^{\infty} \frac{ \left[-i  \omega \left(-1 \right)^m z_1 \right]^{l+r}}{l! r!}  \left(\hat{V} + \hat{V}^\dagger \right)\\
& =  e^{-2 i \omega (-1)^{m} z_1} \left(\hat{V} + \hat{V}^\dagger \right).
\end{align*}
The substitution of the previous equation into \eqref{36} gives us
\begin{align}\label{38}
\left|\psi_{1,2} \right \rangle &=-i e^{-i \omega (-1)^{\hat{n}} z}\int\limits_0^z  e^{-2 i \omega (-1)^{m} z_1} dz_1 \left( \hat{V} + \hat{V}^\dagger \right)  \left|m \right \rangle \nonumber \\
&= -i \frac{  e^{-i \omega (-1)^{m} z} \sin\left[\omega z (-1)^m\right]  }{ \omega (-1)^{m}}   e^{-i \omega (-1)^{\hat{n}} z} \left( \hat{V} + \hat{V}^\dagger \right)  \left|m \right \rangle.
\end{align}
As the sinus is an odd function and $(-1)^{\hat{n}} ( \hat{V} + \hat{V}^\dagger )= ( \hat{V} + \hat{V}^\dagger)(-1)^{\hat{n}+1}$, we arrive to the solution
\begin{equation} \label{39}
\left|\psi_{1,2} \right \rangle = -i \frac{\sin(\omega z)}{ \omega} \left( \hat{V} + \hat{V}^\dagger \right) \left|m \right \rangle.
\end{equation}
Thus, we can build the normalized first-order solution as
\begin{equation} \label{40}
\left|\Psi(z) \right \rangle =  e^{-i \omega (-1)^{m} z} N_1 \left|m \right \rangle -i \frac{\alpha\sin(\omega z)}{\omega}  N_1 \left( \hat{V} + \hat{V}^\dagger \right)\left|m \right \rangle.
\end{equation}
The normalization constant $N_1$ is obtained by considering $k=1$ in Eq.$\eqref{13} $,
\begin{equation} \label{41}
N_1=\left\lbrace  1+ 2 \alpha \Re \left( \left\langle \psi^{(0)} | \psi_{1,2} \right\rangle \right) + \alpha^2 \left\langle \psi_{1,2} | \psi_{1,2} \right\rangle \right\rbrace^{-1/2};
\end{equation}
in this problem the odd powers of perturbative parameter $\alpha$ do not contribute to the normalization constant since $\left\langle m \right|( \hat{V} + \hat{V}^\dagger )^{2n+1} \left|m \right \rangle=0$. Thus, the inner product of $\left|\psi_{1,2} \right \rangle$ with itself is
\begin{equation} \label{42}
\left\langle \psi_{1,2} | \psi_{1,2} \right\rangle = \left[ \frac{\sin(\omega z)}{ \omega} \right]^2   \left\langle m \right|( \hat{V} + \hat{V}^\dagger )^2 \left|m \right \rangle= 2 \left[ \frac{\sin(\omega z)}{ \omega} \right]^2.
\end{equation}
The second-order correction can be obtained again from Eq.$\eqref{10}$,
\begin{equation}\label{43}
\left|\Psi(t) \right \rangle = N_2 \left| \psi^{(0)} \right \rangle + \alpha N_2 \left| \psi_{1,2} \right \rangle + \alpha^2 N_2 \left| \psi_{1,3} \right \rangle.
\end{equation}
Through the application of Eq.$\eqref{9}$, we compute the second-order term $\left| \psi_{1,3} \right \rangle$ as
\begin{equation}\label{44}
\left|\psi_{1,3} \right \rangle = - \frac{ e^{-i \omega (-1)^{m} z}}{\omega} \int\limits_0^z e^{ i \omega (-1)^{m} z_1} \sin(\omega z_1)  d{z_1} \left( \hat{V} + \hat{V}^\dagger \right)^2 \left|m \right \rangle.
\end{equation}
It is easy to see that
\begin{equation}\label{45}
\int\limits_0^z e^{ i \omega (-1)^{m} z_1} \sin(\omega z_1) d{z_1} = \frac{1-\cos(2 \omega z)}{4\omega} + i (-1)^m \left[  \frac{z}{2}-\frac{\sin(2 \omega z)}{4\omega}  \right] ;
\end{equation}
substituting this last result into Eq. \eqref{44} and after some algebra, we get
\begin{equation}\label{46}
\left|\psi_{1,3} \right \rangle = i \frac{ (-1)^m \cos(\omega z)}{2 \omega^2} A(z) \left( \hat{V} + \hat{V}^\dagger \right)^2 \left|m \right \rangle,
\end{equation}
with
\begin{equation}\label{47}
A(z)=\tan \left( \omega z \right)  \left[1+ iz\omega (-1)^m  \right]-z\omega.
\end{equation}
Thus, the solution of wavefunction to second-order is
\begin{align} \label{48}
\left|\Psi(z) \right \rangle=&  e^{-i \omega (-1)^m z} N_2 \left|m \right \rangle - i \frac{\alpha\sin(\omega z) }{\omega}\left( \hat{V} + \hat{V}^\dagger \right) N_2 \left|m \right \rangle \nonumber \\
& + i \frac{ \alpha^2 (-1)^m A(z) \cos(\omega z) }{2 \omega^2} \left( \hat{V} + \hat{V}^\dagger \right)^2   N_2 \left|m \right \rangle.
\end{align}
Using this information and carrying out the sums in Eq.\eqref{13} for $k=2$, the normalization constant is obtained
\begin{align} \label{49}
N_2&=\left\lbrace 1+ \alpha^2 \left[ 2\Re \left( \left\langle \psi^{(0)} |  \psi_{1,3} \right \rangle \right) +\left\langle \psi_{1,2} | \psi_{1,2} \right\rangle \right]+ \alpha^4 \left\langle \psi_{1,3} | \psi_{1,3} \right\rangle  \right\rbrace ^{-1/2} \nonumber \\
&=\left\{ 1+\frac{3}{2} \left[ \frac{\alpha^2 \cos\left( \omega z\right)  \, || A(z)|| }{\omega^2} \right]^2 \right\}^{-1/2},
\end{align}
where the terms that contain $\alpha^2$ sum zero. For the next-order correction, we have
\begin{equation}\label{50}
\left|\Psi(t) \right \rangle = N_3 \left| \psi^{(0)} \right \rangle + \alpha N_3 \left| \psi_{1,2} \right \rangle + \alpha^2 N_3 \left| \psi_{1,3} \right \rangle + \alpha^3 N_3 \left| \psi_{1,4} \right \rangle.
\end{equation}
We get the third-order term from Eq.\eqref{9},
\begin{align} \label{51}
\left|\psi_{1,4} \right \rangle & = \frac{(-1)^m}{2\omega^2} e^{-i \omega (-1)^{\hat{n}} z} \int\limits_0^z  e^{ i \omega (-1)^{\hat{n}} z_1} \cos(\omega z_1) A(z_1)\left( \hat{V} + \hat{V}^\dagger \right)^3 \left|m \right \rangle d{z_1} \nonumber  \\
&=  \frac{i (-1)^m}{2\omega^2} e^{-i \omega (-1)^{m} z} \int\limits_0^z e^{-i \omega (-1)^{m} z_1} \sin(\omega z_1) d{z_1} \left( \hat{V} + \hat{V}^\dagger \right)^3 \left|m \right \rangle \nonumber \\
& \quad - \frac{i (-1)^m}{2\omega} e^{-i \omega (-1)^{m} z}  \int\limits_0^z  z_1 e^{ -2 i \omega (-1)^{m} z_1} d{z_1} \left(\hat{V} + \hat{V}^\dagger \right)^3 \left|m \right \rangle ,
\end{align}
where we have used that $\cos(\omega z_1) A(z_1)=\sin(\omega z_1)-z_1\omega e^{-i \omega(-1)^m z_1} $ and that $(-1)^n ( \hat{V} + \hat{V}^\dagger )^3= ( \hat{V} + \hat{V}^\dagger )^3 (-1)^{\hat{n}+1} $. The evaluation of the integrals via integration by parts leads to
\begin{align}  \label{52}
\left|\psi_{1,4} \right \rangle &=\frac{i \cos(\omega z)}{2 \omega^3} \left[\tan(\omega z)-z\omega \right]= \frac{i B(z)}{2 \omega^3};
\end{align}
thus, \eqref{50} becomes
\begin{align} \label{53}
\left|\Psi(z) \right \rangle& = e^{-i\omega (-1)^m z} N_3 \left|m \right \rangle - i \frac{\alpha \sin(\omega z) }{\omega} \left(\hat{V} + \hat{V}^\dagger \right) N_3 \left|m \right \rangle \nonumber \\
& -  \frac{\alpha^2 }{2 \omega^2} (-1)^m  \left[ z\omega (-1)^m \sin(\omega z)  - i B(z)\right] \left( \hat{V} + \hat{V}^\dagger \right)^2 N_3 \left|m\right \rangle + \frac{ i\alpha^3}{2 \omega^3} B(z)\left( \hat{V} + \hat{V}^\dagger \right)^3 N_3 \left|m \right \rangle.
 \end{align}
The application of the ladder operators to the initial state $\left|m \right \rangle$ gives us the result previously presented in the Section 4. Finally, the normalization constant to this order is given by Eq.\eqref{13},
\begin{equation} \label{54}
N_3=\left\lbrace 1 + \alpha^4 \left[ 2 \Re\left( \left\langle  \psi_{1,2} |  \psi_{1,4} \right \rangle \right)  + \left\langle \psi_{1,3} | \psi_{1,3} \right\rangle\right]  + \alpha^6 \left\langle  \psi_{1,4} |  \psi_{1,4} \right \rangle \right\rbrace^{-1/2}.
\end{equation}

\section{\null}
An alternative procedure to obtain the same results without integration schemes consist in writing the completeness relation $\hat{I}= \sum\limits_{k} \left|k^{(0)} \right \rangle \left\langle k^{(0)} \right| $ in terms of the complete orthonormal set of eigenfunctions of the unperturbed Hamiltonian. If we insert this identity operator inside of Eq.\eqref{6}, together with the initial condition $\displaystyle \left| \psi(0) \right \rangle= \left|n^{(0)} \right \rangle $, we arrive to
\begin{align} \label{55}
\left|\psi_{1,2} \right \rangle
& =  -i t  E_n^{(1)} e^{- i E^{(0)}_n t } \left|n^{(0)} \right \rangle \nonumber \\
&- 2i  \sum\limits_{k \not = n}  e^{- i \frac{t}{2} \left( E^{(0)}_{n} + E^{(0)}_k \right)} \frac{ \sin \left[ \frac{t}{2} \left( E^{(0)}_{n}- E^{(0)}_{k} \right)\right]} { {E^{(0)}_{n} - E^{(0)}_{k}}}  \left\langle k^{(0)} \right|\hat{H}_{p}\left|n^{(0)} \right \rangle  \left|k^{(0)} \right \rangle ,
\end{align}
where the first-order is now expressed in terms of the eigenvalues of $\hat{H_{0}}$,  including also the first-order energy correction written in the form
\begin{equation} \label{56}
 E_n^{(1)}=\left\langle n^{(0)} \right|\hat{H}_{p}\left|n^{(0)} \right \rangle.
\end{equation}
The inner products in \eqref{41} are easy to calculate,
\begin{align} \label{57}
\left\langle \psi_{1,2} |  \psi_{1,2} \right \rangle &= t^2  E^{2(1)}_n + 4 \sum\limits_{k \not = n} \frac{\sin^2\left[ \frac{t}{2} \left(  E^{(0)}_{n}- E^{(0)}_{k} \right) \right]}{\left( E^{(0)}_{n}- E^{(0)}_k \right)^2}\ \left\Vert H_{p_{kn}} \right\Vert^2 ,\nonumber \\
\Re \left( \left\langle \psi^{(0)} |  \psi_{1,2} \right \rangle \right)& =0 .
\end{align}
In this case the normalization constant is given by
\begin{equation} \label{58}
N_1=\left(1+ \lambda^2 \left\langle \psi_{1,2} | \psi_{1,2} \right\rangle   \right)^{-\frac{1}{2}}.
\end{equation}
The derivation for the second order is straightforward, but tedious, and gives
\begin{align}  \label{59}
\left|\psi_{1,3} \right \rangle &= -i e^{-i\hat{H_0} t} \int\limits_0^{t} e^{-i\hat{H}_{0} t_{1}}  \hat{H}_{p} \Big\lbrace -i t_{1}  E_n^{(1)} e^{- i E^{(0)}_n t_{1} }\left|n^{(0)} \right \rangle \nonumber \\
& - 2i \lambda \sum\limits_{k \not = n}  e^{ - i \frac{t_{1}}{2}  \left( E^{(0)}_{k} + E^{(0)}_n \right)} \frac{\sin \left[ \frac{t_{1}}{2}\left( E^{(0)}_{n}- E^{(0)}_k \right)\right]}{E^{(0)}_{n} - E^{(0)}_{k}} H_{p_{kn}} \left|k^{(0)} \right \rangle \Big\rbrace dt_{1},
\end{align}
where the expression inside of the curly brackets is the first order correction. Employing again the identity operator $\hat{I}$ and after some algebraic manipulation, one gets
\begin{align}  \label{60}
\left|\psi_{1,3} \right \rangle =& - e^{- i E^{(0)}_n t } \left( \frac{t^2}{2} E^{2(1)}_n 
+ i t E_n^{(2)} \right) \left|n^{(0)} \right \rangle \nonumber \\
&  + i t e^{- i E^{(0)}_n t } \sum\limits_{k \not = n} \frac{e^{-i \frac{t}{2} \left( E^{(0)}_{k} -E^{(0)}_n \right) } }{E^{(0)}_{n} -E^{(0)}_k} H_{p_{kk}} H_{p_{kn}} \left|k^{(0)} \right \rangle \nonumber\\
& - i t e^{- i E^{(0)}_n t }  E_n^{(1)}  \sum\limits_{k \not = n} \frac{ H_{p_{kn}}}{E^{(0)}_{n} -E^{(0)}_k} \left|k^{(0)} \right \rangle \nonumber \\
& - i e^{- i E^{(0)}_n t }  E_n^{(1)}  \sum\limits_{k \not = n} e^{-i \frac{t}{2} \left( E^{(0)}_{k} -E^{(0)}_n \right) } \frac{\sin\left[\frac{t}{2}\left( E^{(0)}_{k} -E^{(0)}_n \right)\right] }{ \left(  E^{(0)}_{n} -E^{(0)}_k \right)^2}  H_{p_{kn}} \left|k^{(0)} \right \rangle \nonumber\\
& - 2i e^{ -i \frac{t}{2} E^{(0)}_n } \sum\limits_{k \not = n} \sum\limits_{q \not = n} e^{-i \frac{t}{2}  E^{(0)}_{q} } \frac{\sin\left[\frac{t}{2}\left( E^{(0)}_{q} -E^{(0)}_n \right)\right] }{({ E^{(0)}_{q} -E^{(0)}_{n}}) ( {E^{(0)}_{n} -E^{(0)}_{k}})}  H_{p_{qk}} H_{p_{kn}}  \left|q^{(0)} \right \rangle \nonumber\\
&+ 2i \sum\limits_{k \not = n} \sum\limits_{q \not = k}  e^{-i \frac{t}{2} \left( E^{(0)}_{q} + E^{(0)}_k \right) } \frac{\sin\left[\frac{t}{2}\left( E^{(0)}_{q} -E^{(0)}_k \right)\right] }{(E^{(0)}_{q} -E^{(0)}_k) (E^{(0)}_{n} -E^{(0)}_k)}  H_{p_{qk}} H_{p_{kn}}  \left|q^{(0)} \right \rangle
\end{align}
with
\begin{equation} \label{61}
 E_n^{(2)}=\sum\limits_{k \not = n} \frac{|H_{p_{k n}}|^2}{E^{(0)}_{n} -E^{(0)}_k}.
\end{equation}
Considering $ k=2 $ in Eq.\eqref{13},
\begin{align}  \label{59}
 N_2=&\left\lbrace 1 + \lambda^2 \left[2 \Re \left( \left\langle \psi^{(0)} | \psi_{1,3} \right\rangle \right)+ \left\langle \psi_{1,2} | \psi_{1,2} \right\rangle \right]
 \nonumber  \right.  \\ & \left.
 +2 \lambda^3 \Re \left( \left\langle \psi_{1,2} | \psi_{1,3} \right\rangle \right) + \lambda^4 \left\langle \psi_{1,3} | \psi_{1,3} \right\rangle \right\rbrace^{-\frac{1}{2}};
\end{align}
doing the inner products, one can find
\begin{align} \label{60}
&2 \Re \left( \left\langle \psi^{(0)} | \psi_{1,3} \right\rangle \right)= - \left\langle \psi_{1,2} | \psi_{1,2} \right\rangle, \nonumber \\
&2 \Re \left( \left\langle \psi_{1,2}  | \psi_{1,3} \right\rangle \right) = 2t^2 E_n^{(1)} E_n^{(2)}-4 t E_n^{(1)} \sum\limits_{k \not = n} \frac{ \sin^2 \left[ \frac{t}{2} \left( E^{(0)}_{k}- E^{(0)}_n \right)\right]}{ \left(  E^{(0)}_{n} - E^{(0)}_{k}\right)^3} |H_{p_{kn}}|^2 \nonumber\\
&\qquad +4t  \sum\limits_{k \not = n} \frac{ \cos\left[ \frac{t}{2}  \left( E^{(0)}_{k}- E^{(0)}_n \right)\right]\sin \left[ \frac{t}{2} \left( E^{(0)}_{k}- E^{(0)}_n \right)\right]}{ \left(  E^{(0)}_{n} - E^{(0)}_{k}\right)^2} |H_{p_{kn}}|^2 H_{p_{kk}} \nonumber\\
&\qquad + 4  \sum\limits_{k \not = n} \sum\limits_{m \not = n} \frac{ \sin^2 \left[ \frac{t}{2} \left( E^{(0)}_{m}- E^{(0)}_n \right)\right]}{ \left(  E^{(0)}_{n} - E^{(0)}_{m}\right)^2 \left(  E^{(0)}_{n} - E^{(0)}_{k}\right)}H_{p_{nm}} H_{p_{mk}} H_{p_{kn}} \nonumber\\
&\qquad+ 4 \sum\limits_{k \not = n} \sum\limits_{m \not = n} \frac{ \cos\left[ \frac{t}{2}  \left( E^{(0)}_{n}- E^{(0)}_k \right)\right]\sin \left[ \frac{t}{2} \left( E^{(0)}_{m}- E^{(0)}_n \right)\right]}{ \left(  E^{(0)}_{n} - E^{(0)}_{k}\right)\left(  E^{(0)}_{n} - E^{(0)}_{m}\right)\left(  E^{(0)}_{m} - E^{(0)}_{k}\right)} \nonumber \\
& \qquad \qquad \times \sin \left[ \frac{t}{2}  \left( E^{(0)}_{m}- E^{(0)}_k \right)\right] H_{p_{nm}} H_{p_{mk}} H_{p_{kn}}.
\end{align}
Therefore, the normalization constant is
\begin{equation} \label{61}
 N_2=\left[ 1 +\lambda^3 \Re \left( \left\langle \psi_{1,2} | \psi_{1,3} \right\rangle \right) + \lambda^4 \left\langle \psi_{1,3} | \psi_{1,3} \right\rangle \right]^{-\frac{1}{2}}.
\end{equation}

\begin{thebibliography}{99}

\bibitem{1}
\textsf{Griffiths, R.B. J. Stat. Phys. 1984, 36, 219-272.}
 
\bibitem{2}
\textsf{Hartle, J. B.; Hawking, S. W. Phys. Rev. D 1983, 28(12), 2960.}

\bibitem{3}
\textsf{Schr\"{o}dinger, E. Phys. Rev. 1926, 28, 1049. }

\bibitem{4}
\textsf{Rieth, W.; Schommers, W.; Baskoutas, S. Int. J. Mod. Phys. B 2002, 16, 4081-4092. }

\bibitem{5}
\textsf{Ikot, A. N., et al. Journal of Vectorial Relativity 2011, 6, 65-76.}

\bibitem{6}
\textsf{Ray, J. R. Phys. Rev. A 1982, 26, 729.}

\bibitem{7}
\textsf{Burgan, J. R.; Feix, M. R.; Fijalkow, E.; Munier, A. Phys. Rev. A 1979, 74(1-2), 11-14.}

\bibitem{8}
\textsf{Khandekar, D. C.; Lawande, S. V.; J. Math. Phys. 1979, 20, 1870-1877. }

\bibitem{9}
\textsf{Schr\"{o}dinger, E. Annalen der physik, 1926, 385, 437-490.}

\bibitem{10}
\textsf{Strauss, H. L. \emph{Quantum mechanics: An Introduction}; Prentice-Hall, New Jersey, 1968. }

\bibitem{11}
\textsf{Lamb, W. E. Phys. Rev. 1952, 85, 259.}

\bibitem{12}
\textsf{Sakurai, \emph{J. J. Modern quantum mechanics, revised edition}; Addison-Wesley, M A, 1995.}

\bibitem{13}
\textsf{Møller, C.; Plesset, M.S. Phys. Rev. 1934, 46, 618.}

\bibitem{14}
\textsf{Simon, B.; Dicke, A. Ann. Phys. 1970, 58, 76-136.}

\bibitem{15}
\textsf{Rother, T.; J. Electromagn. Waves Appl. 1993, 7, 857-871.}

\bibitem{17}
\textsf{Mart\'inez-Carranza, J.; Soto-Eguibar, F.; Moya-Cessa, H. Eur. Phys. J. D. 2012, 66(1), 1-6.}

\bibitem{16}
\textsf{Mart\'inez-Carranza, J.; Moya-Cessa, H.; Soto-Eguibar, F. \emph{La teor\'ia de perturbaciones en la mec\'anica cu\'antica}; Editorial Acad\'emica Espa\~nola, 2012.}

\bibitem{18}
\textsf{Villegas-Mart\'inez, B. M.; Soto-Eguibar, F.; Moya-Cessa,  \emph{Application of Perturbation Theory to a Master Equation};, Adv. Math. Phys. 2016, 9265039, 7, 2016.doi:10.1155/2016/9265039 }


\bibitem{19}
\textsf{Fetter, A. L.; Walecka, J. D.\emph{ Quantum theory of many-particle systems}; McGraw-Hill, New York, 2012.}

\bibitem{20}
\textsf{Johnson, G. W.; Lapidus, M. L. \emph{The Feynman integral and Feynman's operational calculus}; Oxford Univ. Press, Oxford and New York, 2000.}

\bibitem{21}
\textsf{Peschel, U.; Pertsch, T.; Lederer, F. Opt. Lett. 1998, 23, 1701-1703.}

\bibitem{22}
\textsf{Ablowitz, M.J.; Musslimani, Z.H. Physica D 2003, 184, 276--303.}

\bibitem{23}
\textsf{Poulios, Konstantinos, et al. Phys. Rev.Lett. 2014, 112, 143604.}

\bibitem{Leija} \textsf{Weimann, S.;  et al. Nature Comm. 2016,  7, 11027.}

\bibitem{24}
\textsf{Longhi, S. Opt. Lett. 2010, 35, 235-237.}

\bibitem{25}
\textsf{Moya-Cessa, H.M.; Soto-Eguibar, F. \emph{ Differential Equations:  An operational approach}; Rinton Press, New Jersey, 2011. }

\bibitem{Kenkre} 
\textsf{Kovanis, V.I.; Kenkre, V.M. Physics Letters A 1988, 130, 147.}

\bibitem{26}
\textsf{Soto-Eguibar, F.; Moya-Cessa, H.M. Int. J. Quantum Inf. 2012, 10, 1250072.}

\bibitem{27}
\textsf{Klimov, A.; Sanchez-Soto, L.L. Phys. Rev. A 2000, 61, 063802.}

\end{thebibliography}
\end{document}